# Criteria for the growth of fullerenes and single-walled carbon nanotubes in sooting environments

**Shoaib Ahmad**

*National Centre for Physics, Quaid-i-Azam University Campus, Shahdara Valley, Islamabad, 44000, Pakistan*

Email: sahmad.ncp@gmail.com

**Abstract**

The spherical curvature induced by pentagons in corannulenes and hexagonal sheets is shown to be the basic constituent that controls the growth of fullerenes and single-walled carbon nanotubes (SWNTs) in soot forming and carbon vapour environments. Formation of the initial ring of five or six atoms is the essential step which with the addition of further pentagons and hexagons determines whether a spinning fullerene is to be formed or the cap that lifts up and leads to the formation of an SWNT. A continuum elastic model is developed to determine the criteria for the growth of these structures. The observed dominance of the growth of 14 Å diameter armchair SWNTs in sooting and carbonaceous environments is explained by using the nanoelastic model of C shells.

## 1. Introduction

The growth of closed cages of carbon that include fullerenes and nanotubes has been actively pursued in the last two decades ever since the discoveries of Kroto *et al* [1] and Iijima [2] and a number of review articles and books have been written [3]. The mechanisms by which the cage closure occurs is, however, still debated—as to why one type of structure like fullerene is favoured under certain conditions as opposed to the formation of nanotubes or vice versa. This question rises in particular in those experimental methods like the arc discharge, laser ablation and CVD [3] where both kinds of carbon nanostructures have been produced. Variations in the experimental parameters like the electrode geometry, gas pressure, and temperature of the substrates and the presence of catalysts like Fe, Co, Ni are generally quoted as being the reasons for the preferential growth of fullerenes or nanotubes. The identification of the mechanisms that produce either the fullerenes or the nanotubes still needs to be dealt with in a coherent manner.

The corannulene road for the growth of fullerenes has been proposed using topological arguments [4] but the physics and the basic mechanisms of the importance of the curvature that is inherent in the pentagon-centred corannulene have been lacking. In this communication, we propose a mechanism by which cage closure is explained and the criteria that govern the growth of fullerenes and single-walled nanotubes in sooting environments are deduced. An earlier elastic continuum model for the shelled structures of carbon [5] is used to prove that the curvature-related growth of the fullerenes and nanotubes can be associated with the formation of the initial bowl of corannulene or the addition of pentagons in an all-hexagon sheet. Corannulene, however, is shown to be the basic structural unit in the self-assembly of single-walled carbon nanotubes while fullerenes can grow in both situations, in which either a pentagon or hexagon is the first stable ring to be formed. It is shown that whereas C60 is the most dominant species among the fullerenes produced under optimum conditions in carbonaceous discharges, the growth of a SWNT around a hemispherical cap of C240 in the armchair configuration is the preferential nanotube [6, 7] with diameter ∼14 Å. 70–90% of SWNTs grown in a catalyst-free environment by laser ablation and carbon arc discharge techniques [8] also have an optimized nanotube diameter of 14 Å. From the continuum elastic considerations of the curvature introduced by the six corannulenes that are necessary for the cap of any SWNT, the condition of non-abutting corannulenes emerges as the most essential condition for the dominant growth of 14 Å diameter SWNTs.

Our model describes the elasticity of carbon nanostructures and one of the major conclusions drawn is that the fullerenes when formed will be spinning. $C_{60}$ having the perfect symmetry with non-abutting pentagons and all-abutting corannulenes has been observed to be the ideal structure to grow in various kinds of sooting environments. The fullerenes are formed from either the five- or six-member C rings; the growth mechanisms and various geometrical configurations are discussed in sections 5 and 6. Initiation of embryos around a heptagon is not ruled out in this analysis; neither is the possibility of its addition at a later stage. The analysis based on the curvature induced by pentagons in corannulenes can be extended to heptagonal curvatures as well. The free-standing, single-walled carbon nanotubes, however, are shown to be the direct product of the overgrowth of the initial corannulenes.

Non-abutting corannulenes has been shown to be an essential condition for the growth of an SWNT from a sooted surface by the addition of $C_2$s. At the same time, the model presented here does not rule out the possibilities of the growth of fullerenes that are smaller or larger than $C_{60}$, nor have SWNTs with diameters other than 14 Å been excluded. In addition, multi-shelled structures like the carbon onions and multi-walled nanotubes (MWNTs) can also be described by extending the same model. In this paper, the optimum conditions for the growth of $C_{60}$ and 14 Å diameter SWNTs in catalyst-free carbon vapour have been derived; the extension of the model to non-equilibrium environments can be attempted using similar methods.

## 2. Model for the nanoelasticity of C shells

Fullerenes, being nano-spheroidal C cages, and nanotubes, as composite structures with half-fullerenes capped on cylindrically bent graphene, have been studied utilizing elasticity theory by various authors [9–17]. In the model presented here, curvature-related conditions are derived for describing the growth of fullerenes and nanotubes. The spherical curvature inherent in the nascent $sp^2$ bonded corannulene and the curvature introduced by the addition of pentagons in hexagonal sheets is described as the fundamental requirement for the growth of any shelled structure. This curvature is explained from the shell theory of C nanoparticles and as a result this approach is characteristically different from that based on the bending of graphene sheets to produce spherical curvature [9–15]. In developing the nanoelastic model of C-shells it is clarified that fullerenes or their respective half-caps may not be comparable to the bent graphene sheets. This is because the elastic behaviour of shells versus plates is manifested in the stretching and bending effects which appear in reverse order for the two kinds of structures.

Existence of spherically curved surfaces introduces additional forces that depend on the properties of the particular surfaces. The coefficient of surface tension $T$ is related, in the case of the separation of phases, to the pressure difference $\Delta P$ across the interface by Laplace's equation $\Delta P = 2T/R$, where $R$ is the radius of curvature. Elasticity theory relates $T$ to the tangential stresses. In the case of a C nano-shell, the equilibrium equation requires the calculation of these stress tensors. Any deviations from sphericity are dealt with by taking stretching as the first-order and bending as the second-order effect. On the other hand, the conversion of a graphene sheet into a cylindrical tubule does not require the stretching effect to be considered; bending is the first-order effect. The strain energies obtained for tubules with the consideration of bending only [9] can and do provide fairly accurate estimates, whereas the evaluation of the deformation energies that are required for the pentagonal protrusions in shelled structures requires the stretching and bending effects to be taken into account together.

The following basic assumptions have been made to establish the nanoelasticity of C shells:
(1) The fullerene shell is assumed to be equivalent to an appropriate Goldberg polyhedron with 12 pentagons and a variable number of hexagons [18].
(2) The thickness $t$ of the shell is expected to remain constant even during deformations of the shell. In this analysis $t$ is treated as the thickness of a fullerene shell which for $C_{60}$ with radius 3.52 Å yields an average C atom diameter of 1.82 Å. This value of $t$ is used in this communication.

(3) The elastic modulus $Y = 10^{12}$ Pa and Poisson ratio $v = 0.163$ remain constant in fullerenes and the cap nanotubes of different radii; we use the average numerical values for the basal plane of graphite from [19].

(4) When dealing with pentagonal deformations, the stretching and bending effects are to be considered together.

The expected achievements of such a model are the explanations that it will offer for: (1) the importance of the role of the spherical curvature induced by pentagons in the embryonic structures, (2) the identification of the conditions that lead to the formation of either a fullerene or a nanotube, (3) the explanation for the dominance of $C_{60}$ among the fullerenes and (4) the reason for the observations of the dominance of 14 Å diameter armchair nanotubes, i.e. a SWNT with a half-$C_{240}$ as the cap in sooting and carbonaceous environments.

## 3. The corannulene

Formation of the initial pentagon surrounded by five hexagons produces a corannulene that can serve as the role model for the curved structures as depicted in figure 1.

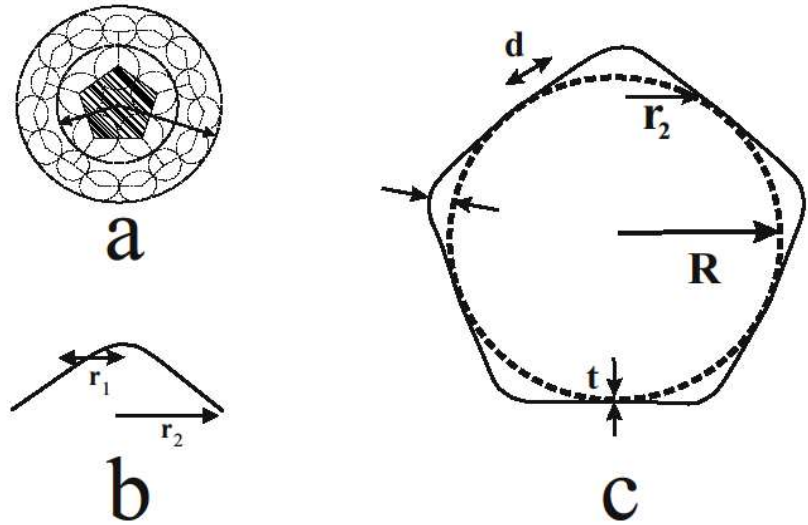

**Figure 1.** (a) The pentagon-containing corannulene is shown with two effective dimensions shown by arrows. (b) The same 20-atom structure (1 pentagon + 5 hexagons) of the corannulene is shown tangentially to reveal the role that $r_1$ and $r_2$ play in the formation of shells. (c) A typical large shell is shown with outward protruding corannulenes; all the relevant parameters are also shown. The outward protruding structure is superimposed on an inner sphere of radius $R$, thickness $t$; the protrusion is shown to have dimension $\zeta$ which is the difference between the radii of the circumscribing and inscribing spheres. The corannulene caps $\sim r_2$ are along the meridian with dimension $d \sim \sqrt{t R}$. These identify the bulge.

However, it must be pointed out that the curvature appears only after the fifth hexagon has been added to the corannulene. In figures 1(a) and (b) the two effective dimensions are shown as $r_1$ and $r_2$, that define the central region of the atoms forming the pentagon ($r_1 \sim 2$ Å) and the corannulene containing 20 C atoms ($r_2 \sim 4$ Å), respectively. The values of $r_1$ and $r_2$ are obtained from the average C–C bond lengths $d_{C-C}$ cited in the literature [3, 16] with $d_{C-C} = 1.45$ Å in pentagons and $d_{C-C} = 1.40$ Å for hexagons; the respective areas are 4.37 and 5.25 Å$^2$. The two radial dimensions $r_1$ and $r_2$ play an important role in determining the resultant stresses that are generated and become pivotal to the formation of either the shelled fullerene or the hemispherical cap that lifts the nanotube underneath. Figure 1(c) can be viewed as either a complete fullerene shell with 12 protruding corannulenes or as a hemispherical cap centred at a pentagon along $C_5$ axis with five circumferential corannulenes. It could also represent a large fullerene (e.g., $C_{240}$ or $C_{540}$) or the top cap of a single-walled nanotube (SWNT). It shows outward protruding structures superimposed on an inner shell of radius $R$ and thickness $t$. The protrusion has a dimension $\zeta$ that can be calculated by assuming that the shell (or half-shell) is equivalent to an appropriate Goldberg polyhedron and $\zeta$ is the difference in radii of the circumscribing and inscribing spheres (or circles in the case of the cap). Figure 1(c) shows another important dimension $d \sim \sqrt{tR}$ which is the circular strip at the edges of the outward bulges in which a major part of the elastic energy is stored.

When hexagons are added around the initial lifting pentagons as shown in figure 1, the resulting structures start to exhibit two interlinked curvatures; one of these is defined by the inter-pentagonal distance, $r_2$, and the other is that of the pentagonal protrusion itself, i.e., $r_1$. Therefore, the existence of the pentagonal protrusion in a shell-like structure is equivalent to curvature in the curved surface. $C_{240}$ is a case in point where each pentagon is surrounded by a corannulene. The resulting spherical shell has 12 pentagonal protrusions symmetrically embedded on a sphere of radius 7.05 Å. Even if structures other than fullerenes are to form in a carbonaceous vapour, these pentagons and their associated outward bulging surfaces provide curvature in the curved surfaces. The emergence of a bulge, therefore, indicates the existence of an internal force whose magnitude can be determined from the size and extent of the bulge. For the present investigations we will first estimate the extent of the pentagonal protrusion $\zeta$ superimposed on an underlying spherical surface and provide order of magnitude estimates of this outward lifting force $f_o$.

## 4. The outward force $f_o$ and the critical stress $P_{cr}$

When a spherical shell of thickness $t$ is subjected to a concentrated force per unit area $f_o$ along the outward normal, the resulting deformation $\zeta = R' - R$, where $R$ and $R'$ are the radii of the inscribing and circumscribing spheres, respectively. One has to determine the size of the protrusion $\zeta$ as a function of the outward force $f_o$. The major part of the elastic energy is stored in the narrow bending strip $\sim d$ around the edge of the bulge. Geometrically, angle $\alpha$ is the angle subtended by the bulge at the spherical centre and is $= \zeta/d = r_2/R$; the area of this strip, $\sim r_2 d\zeta$, will vary over a perpendicular distance $d$. The inscribing shell radius $R$ can be considered to be equivalent to the radius of the corresponding fullerene. The distance $d$ is a measure of the region of outward protrusion along the pentagon and the resulting area of deformation $\sim d^2$. The deformation energies around this region can be estimated [5, 20, 21] as the bending energy over this area

$$E_{ben} \sim Y t^3 (\zeta/d)^2,$$

and the respective stretching energy

$$E_{str} \sim Y t [(\zeta d)/R]^2.$$

The bending energy decreases and the stretching increases with increase in $d$; thus both energies should be considered in determining the deformation near the point of application of the force. Minimizing their sum ($E_{ben} + E_{str}$), one gets $d \sim \sqrt{t} R$. As can be visualized from figure 1(c), stretching is along the meridian and bending along the circle of latitude (of radius $r_2$).

The total elastic energy in the bending strip of a corannulene is $E_{cor}^{ben} \approx Y t^{5/2} (\zeta^{3/2}/R)$. The total stretching energy in the bulge is $E_{cor}^{str} \approx Y (t/R)^2 (tr)$, where $r = r_2$. For a given value of the deformation one can obtain the outward force $f_o$ by equating it to the derivative of $E_{cor}^{ben}$ with respect to $\zeta$ as

$$f_o \approx Y t^{5/2} (\zeta^{1/2}/R) \qquad (1)$$

As the bulge or the outward protrusion has occurred due to the generation of internal stresses associated with the curvature of the corannulene, this stress $P$ can be associated with the work done in producing a defect volume $\sim \Delta V$, where $\Delta V \sim \zeta r_2^2 \sim \zeta^2 R$. The total free energy is $[E_{cor}^{str} - P\zeta^2 R]$. Taking the derivative of this total free energy yields

$$\zeta \approx Y^2 t^5 / (R^4 P^2) \qquad (2)$$

This is an inverse relation between $\zeta$ and $P$ ($\zeta$ increases when $P$ decreases); hence an unstable equilibrium is indicated and, therefore, the bulges with large $\zeta$ grow of their own accord, while the smaller ones shrink. The equation (2) corresponds to a maximum of the total free energy. A critical value of $P$ exists for $\zeta \approx t$ beyond which even small changes in the shape of shells increase spontaneously. This value of the critical P is

$$P_{cr} \approx Y t^{5/2} / (R^2 \zeta^{1/2}) \qquad (3)$$

In figure 2 the upward lifting force per unit area $f_o$ from equation (1) on a corannulene and the critical stress $P_{cr}$ from equation (3) that is generated in the shelled structure are plotted as a function of $C_x$ where $x$ is the number of C atoms in a complete fullerene. The crossover is around $C_{240}$ when the critical pressure for the bulges to grow becomes smaller than the outward bulging force and the structure becomes unstable. Although the figure has been calculated for a complete fullerene with 12 corannulenes, the results are equally valid for a hemispherical cap that will lift and pull an SWNT underneath.

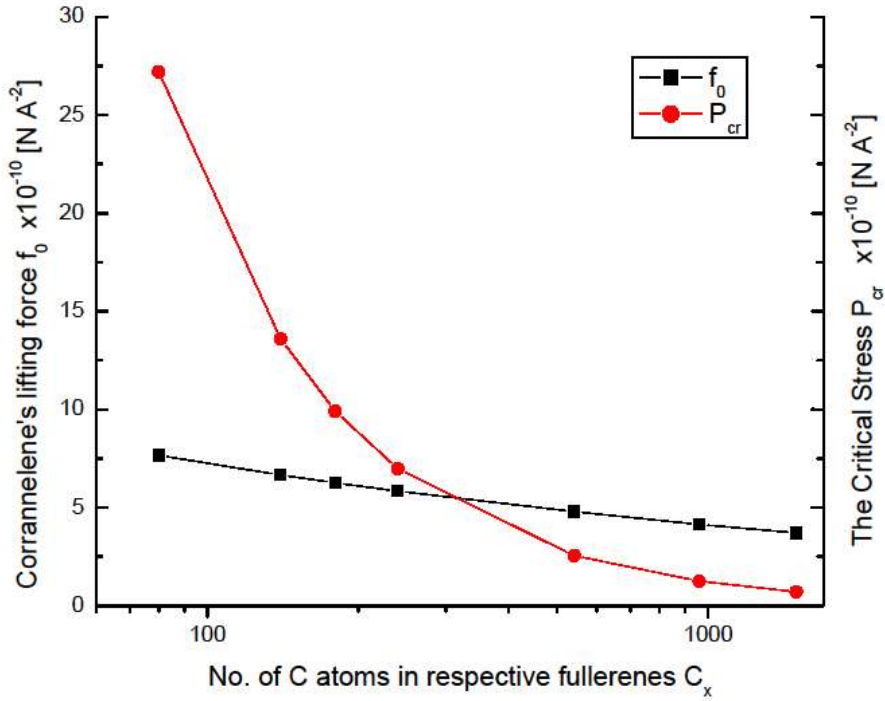

**Figure 2.** The upward lifting force per unit area $f_o$ (equation (1)) on a corannulene and the associated critical stress $P_{cr}$ (equation (3)) are plotted as a function of $C_x$ where $x$ is the number of C atoms in respective fullerenes. The crossover is around $C_{240}$ when the critical stress for the bulges to grow becomes smaller than the outward bulging force and the structure becomes unstable.

## 5. The armchair and zigzag icosahedral caps

The fullerene spheroids when treated as Goldberg polyhedra reveal the symmetries of respective fullerenes around the chosen axes of symmetry. The vertices of these polyhedra are given by $n = 20(b^2 + bc + c^2)$, where $b = c = 1, 2, 3$ for $C_{60}$, $C_{240}$ and $C_{540}$, respectively. These are the only fullerenes that can be shelled inside each other to produce carbon onions [22] with the same mutual spacing as that which exists in graphite layers, i.e., 3.34 Å. The other set of icosahedra, $C_{80}$ and $C_{180}$, have $b = 2, 3$ and $c = 0$, respectively. Figure 3 shows a set of four icosahedral caps that can grow around a central corannulene shown as the dotted circle at the centre. These caps respectively belong to $C_{60}$ (figure 3(a)), $C_{240}$ (figure 3(b)), $C_{80}$ (figure 3(c)) and $C_{180}$ (figure 3(d)). The characteristic feature of these icosahedra is the way in which the six pentagons of the hemispherical caps are geometrically arranged.

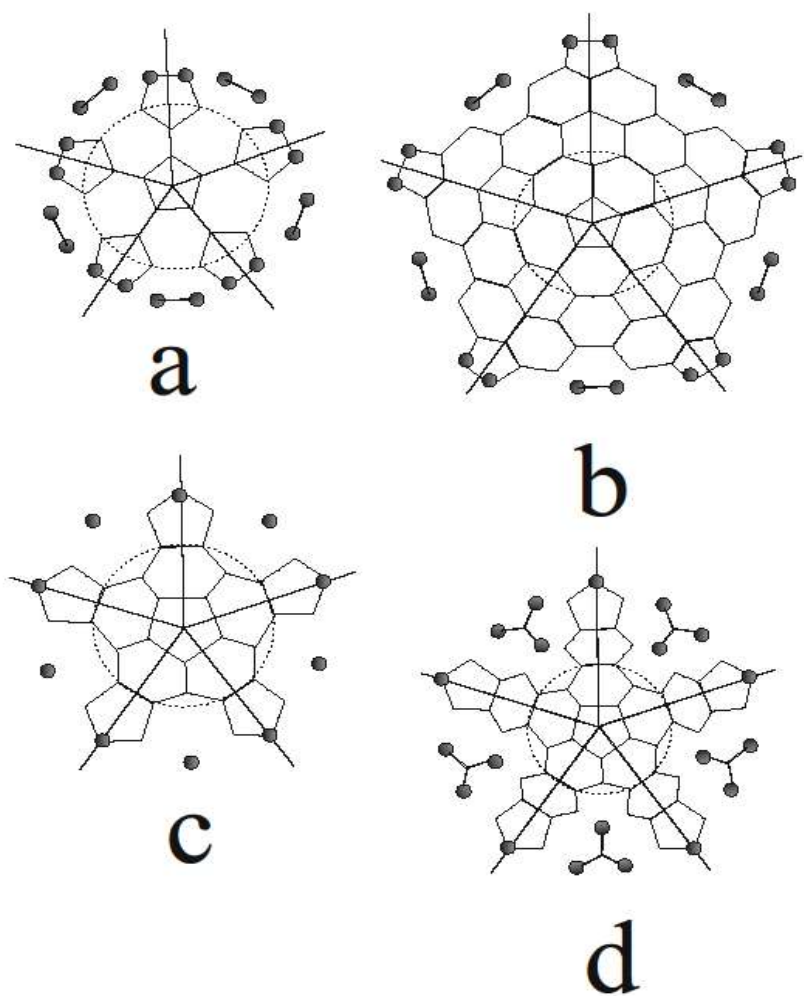

**Figure 3.** A set of four icosahedral caps are shown that can grow around a central corannulene shown as the dotted circle at the centre. These caps respectively belong to $C_{60}$ (a), $C_{240}$ (b), $C_{80}$ (c) and $C_{180}$ (d). The common feature of the four icosahedra is the central pentagon and the circumferential five pentagons in armchair ((a), (b)) and zigzag ((c), (d)) geometries. The filled circles are the C atoms that depict the pattern of growth out of these structures that may either lead to fullerene formation or the growth of a single-walled nanotube.

The central pentagon and the circumferential five pentagons are shown in armchair (figures 3(a) and (b)) and zigzag (figures 3(c) and (d)) geometries around the same $C_5$ axis. The filled circles are the C atoms that depict the pattern of growth out of these structures that may either lead to fullerene formation or the growth of a single-walled nanotube.

Once an armchair cap is formed, it grows further as an SWNT with the addition of C dimers $C_2$. Each $C_2$ is then equivalent to a hexagon that provides the tubule in an armchair configuration. The growth mode would require $C_2$ addition at the roots of the outward growing nanotube. The requirement for the further growth of $C_{80}$ and $C_{180}$ is the addition of $C_1$ and $C_3$, as shown. The growth of the zigzag nanotubes requires a carbonaceous vapour that is equally rich in $C_1$, $C_2$ and $C_3$. Thus an upward growing SWNT in zigzag mode has a barrier in terms of the statistical requirement of almost equal abundance of the three component species of the growth, i.e., $C_1$, $C_2$ and $C_3$, as opposed to armchair SWNTs that need only $C_2$. The four icosahedra presented in figure 4 are shown around a central hexagon along the $C_3$ axis.

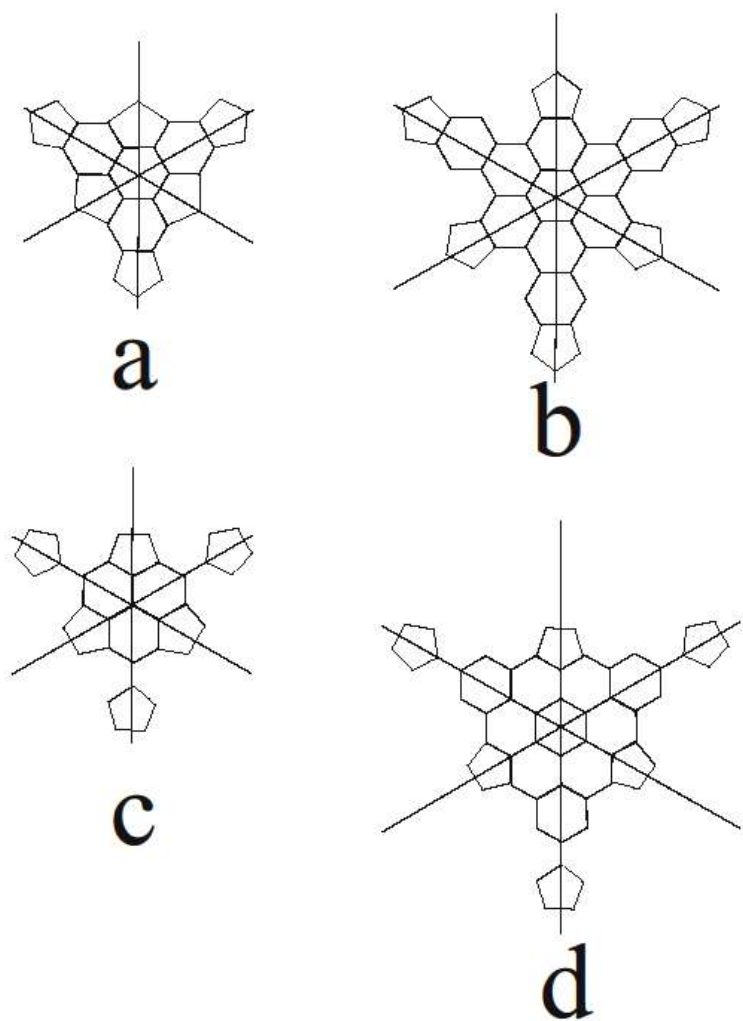

**Figure 4.** The four icosahedra shown in figure 3 are shown again around a central hexagon (with the exception of $C_{80}$) along the $C_3$ axis. In the case of (a) and (b), a full coronene (a flat hexagon-only bulging block of graphene) does not develop while, in the case of (c) and (d), it develops as the centrepiece. It can be seen that along $C_3$ axis the structures belonging to $C_{60}$ and $C_{240}$ produce zigzag geometry while $C_{80}$ and $C_{180}$ produce armchair arrangements.

The common feature of all the four structures is the dispersal of two sets of three pentagons each, around the $C_3$ axis. The peripheral set of three pentagons is in the zigzag configuration for $C_{60}$ (figure 4(a)) and $C_{240}$ (figure 4(b)), while it produces an armchair arrangement for $C_{80}$ (figure 4(c)) and $C_{180}$ (figure 4(d)). Therefore, the requirement (of $C_1$, $C_2$ and $C_3$) for the further growth of icosahedra that are growing around the C3 axis is exactly opposite to that for the two sets shown in figure 3. In the case of figures 4(a) and (c), a full coronene (a flat hexagon building block of graphene consisting of seven hexagons) does not develop while, in the case of figures 4(b) and (d), it develops as the centrepiece.

## 6. Corannulene versus coronene as embryos for the spinning fullerenes or the outward growing SWNTs

If a corannulene has just been formed in a sooting environment then there exists a strong inward bending moment along the circumference $M_0 \sim (1 + v)D/r$ where $D$ is the flexural rigidity of graphene: $D = Y t^3/\{12(1 - v^2)\} \approx 3$ eV or $5.16 \times 10^{-19}$ N m; we have used $Y = 10^{12}$ Pa, $v = 0.163$ and $t = 1.82$ Å. For the corannulene with $r_1 \sim 2$ Å, this bending moment $M_0 \approx 3 \times 10^{-9}$ N. This introduces a bending stress $6M_0/t^2 \sim 3 \times 10^{-9}$ N Å$^{-2}$. If the addition of pentagons takes place then the bending stresses are further assisted in setting the evolving structure into rotations. The torque on the corannulene of moment of inertia $I_0 \sim 2 \times 10^{-45}$ kg m$^2$ produces angular acceleration $\sim 10^{25}$ rad s$^{-2}$. The rotational energy of a 3D rotor is a function of temperature $T$ and the moment of inertia can be written as $E_{rot} = k_B T \times \ln(8\pi 2(I_0 k_B T/2\pi^2 h^2)^{3/2})$ [23]. $E_{rot}$ is also $= 1/2 \, I_0 \omega^2$, $\omega$ is the angular frequency. A rotating structure at 300 K yields a spinning speed of $10^{12}$ s$^{-1}$. A rotating curved structure has a higher probability of accreting $C_1$, $C_2$ and $C_3$ from the sooting environment to complete the spheroid. These structures (from $C_{20}$ ... $C_{60}$ ...) are predicted to be spinning with rotational frequencies $\sim 10^{-12}$ s. Our results may provide the justification for the observations of spinning $C_{60}$ with rotational frequency $\sim 10^{-12}$ s in fullerite [23, 24].

There exist a wide range of values of the parameters $Y$ and $t$ for graphite, fullerenes and nanotubes used by different researchers [25–29]. The range of thickness $t$ used or derived from experiments is between 0.7 and 3.34 Å while those of $Y$ are between 5.5 and 1 TPa. Formation of an initial hexagon also starts the process that can produce shells and nanotubes. Figure 4 describes four such situations where the hexagonal net expands until a pentagon is added and the curvature is introduced with the associated bending moment. Symmetry imposes restrictions on the way pentagons can be added to the hexagonal sheet. The spherical curvature is introduced by the two sets of three pentagons each, to complete the hemispherical shell. Both of these sets are displaced by 60° from each other and symmetrically disposed around the centre of the initial hexagonal net. In the case of $C_{60}$ and $C_{80}$, the hexagonal net comprises of one and three hexagons, respectively, whereas a full coronene (seven hexagons) develops in the case of $C_{180}$ and $C_{240}$ before the addition of pentagons.

This requirement of symmetry imposes a constraint on the hexagon initiated structures to produce the spinning fullerenes as the dominant by-product for C accretion. An SWNT is less likely to grow along the $C_3$ axis due to the rotational torque introduced in two separate stages of three-pentagon addition. Therefore, a closed shell is more likely to be formed shown in figure 4.

Right from the moment of the corannulene formation, the earlier addition of pentagons enhances the probability of fullerene formation. Spinning is the essential step that provides a barrier to the formation of nanotubes and it is overcome when addition of the remaining pentagons is delayed by the formation of a hexagonal network around the central corannulene as in figures 3(b) and (d). In the case of the $C_{240}$ cap shown in figure 3(b), thirty hexagons surround the central pentagon in the form of three rings. The addition of the remaining five pentagons takes place at the periphery. The bending stress introduced and uniformly distributed by the circumferential pentagons is about 7 Å away from the central pentagon and may introduce an upward lift to the cap as opposed to the spin in the case of the smaller ones. The case for the $C_{180}$ cap shown in figure 3(d) is equally interesting and worth considering as a candidate for the preferential growth of a zigzag SWNT with 12 Å diameter and not acting as a seed for the growth of a fullerene for the same reasons. Here again the addition of 30 hexagons around the central pentagon provides the delay in the pentagon-induced bending stresses until the addition at the circumference occurs. In this case the resulting cap is in zigzag geometry and the only difficulty for the growth of this cap is the requirement of addition of $C_1$ and $C_3$ to provide for the necessary tubule formation, as opposed to the requirements of the addition of $C_2$s in the growth of armchair SWNTs.

## 7. Necessity of the non-abutting corannulenes for SWNT

The outward lift introduced by the pentagonal protrusion starts with an initial curvature $r_1 \sim 2$ Å; the addition of hexagons stabilizes it by distributing the stress in a larger area, $\sim 51$ Å$^2$. Fullerenes smaller than $C_{60}$ have adjacent pentagons and have larger curvature-related stresses. The associated strain is not uniformly distributed in the polyhedral structures that do not display icosahedral symmetry. All of the 12 pentagons are abutting in $C_{20}$ and the resulting strain is maximized in the shell of diameter 4 Å. $C_{20}$, therefore, is a highly strained structure and an example of an inherently unstable fullerene due to the pentagon-induced strains, whereas in the case of $C_{60}$, the 12 non-adjacent pentagons form a perfect spherical shell that possesses the unique distinction of sharing all 60 C atoms uniformly in its 12 pentagons, and 20 hexagons with the 12 corannulenes emerge as the by-product of sharing all sets of five atoms producing pentagons in such a way that these are used up to form the surrounding ring of five hexagons. All the corannulenes are abutting in perfect symmetry in $C_{60}$ and the curvature-related strain is uniformly distributed over the spherical shell.

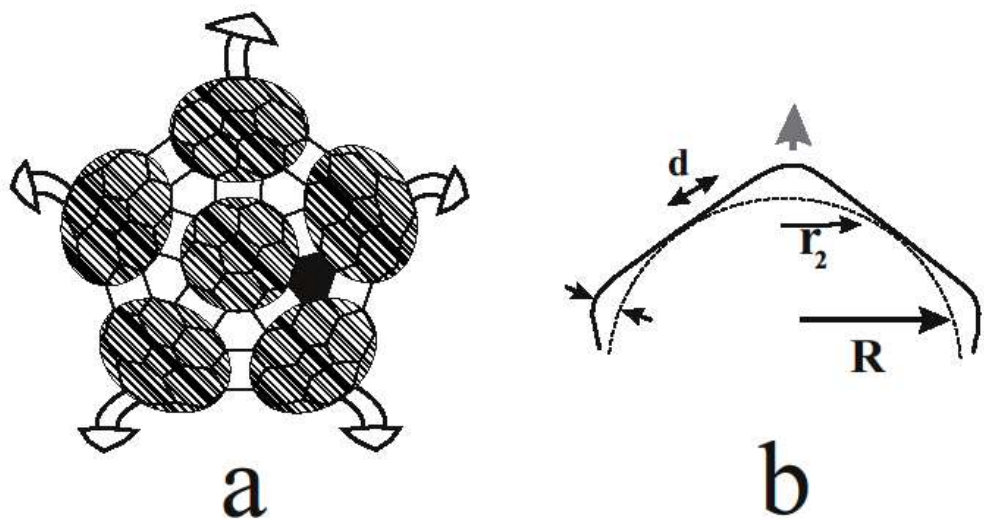

**Figure 5.** (a) The hemispherical cap belonging to $C_{240}$ of diameter 14 Å is shown. Six non-abutting corannulenes are shown as hatched areas; the five arrows along the circumference indicate the bending moments as well as the direction of growth of the structure dependent upon the addition of dimers ($C_2$). The net outcome is the armchair SWNT. A blackened hexagon indicates the option of growth centred around a coronene as shown in figure 4(b). (b) The same structure as in (a) but shown along the direction of upward growth indicated by the heavy arrow. The four typical dimensions are the same as in figure 1(c).

Shelled structures with radii larger than 3.5 Å have interpenetrating corannulenes with increasing separation of pentagons and the resulting bending stresses are distributed in larger areas. The description of the structure growing around a central corannulene in sections 3 and 4 provides a measure for the two regions of the concentration of the stretching and bending energies for the outward protruding pentagonal defects. The outward lift results from the stretching of a region of curvature $r_1$ and area ∼14 Å$^2$. The next stage is the addition of the C atoms from the surrounding C vapour to introduce either pentagons or hexagons around the central, lifting pentagon. The addition of hexagons stabilizes it by distributing the stress in a larger area ∼51 Å$^2$. In figure 5(a) the armchair cap belonging to $C_{240}$ of diameter 14 Å is shown with the six corannulenes as hatched areas. Each corrannulene is surrounded by a ring of 10 hexagons. This is the bending strip shown as region *d* in figure 5(b) in which the bending energy is concentrated. A blackened hexagon indicates the alternative option of growth centred on one of the five possible coronenes. Such geometry is shown in figure 4(b) where a zigzag cap is formed. In figure 5(b) the same structure as in figure 5(a) is shown along the direction of upward growth indicated by the heavy arrow. The four typical dimensions are the same as in figure 1(c). The 14 Å hemispherical caps require 90 C atoms to form 6 pentagons and 30 hexagons as shown in figure 5(a). To complete the half-cap of $C_{240}$ needs 120 C atoms that yield six non-abutting corannulenes. Figure 5(a) shows such an armchair cap belonging to $C_{240}$. These non-abutting corannulenes are shown as hatched areas; the five arrows along the circumference indicate the bending moments as well as the direction of growth of the nascent structure. The net outcome may be an armchair SWNT or a fullerene, depending upon the addition of $C_2$s only or a mixture of $C_1$, $C_2$ and $C_3$, respectively. Cage closure to produce a $C_{240}$ fullerene, however, is less likely when the structure is rising from a sooted surface. It may be more probable when C accretion is isotropic and the fullerene is forming in vapour as opposed to rising out of the sooted surface.

$C_{70}$ provides experimental evidence in favour of the above analysis, where the absence of the non-abutting corannulenes stops the growth of the smallest and the second-most prominent member of the fullerene family as a nanotube of diameter 7 Å and length 9.5 Å. It comprises the two halves of $C_{60}$ with a ring of five hexagons added in an armchair configuration. $C_{70}$ can also be considered the smallest, most stable nanotube that grows in environments that favour the formation of fullerenes. It is also analogous to the classic case of the difficulties of gender identification; the broad definition of fullerenes [18] shows $C_{70}$, as a fullerene in symmetry group $D_{5h}$, as a symmetric top that has evolved out of an icosahedral cap belonging to the spherical top category. The curvature induced by the six abutting corannulenes in the half-cap of $C_{60}$ is stabilized by the addition of five hexagons that share their strain with a similar half-cap to produce a highly symmetric single-walled nanotube where the elastic strains have refused to allow the further growth of the nanotube. Our criterion for the spinning structures provides justification for the dominance of fullerene formation as opposed to that of the nanotubes within the C vapour.

## 8. Conclusions and future directions

The paper presents the results and rationale for the emphasis to be given to the early growth stages of the clusters of the curved C atoms that will form the cap of a nanotube or a fullerene. The above analysis provides a picture of typical soot formation with carbon vapour enclosed in an environment where C accretion can occur to produce planar and curved C embryos for the growth of fullerenes and nanotubes. The growth of the larger 3D structures around the 2D rings of five or six C atoms is dependent upon the elastic properties related to their respective geometries. The spherical curvature and the emerging symmetry of the structure are identified as the basic characteristics that control the growth of fullerenes and single-walled carbon nanotubes (SWNTs) in all sooting environments. Addition of pentagons to an earlier formed hexagon(s) has been shown to lead predominantly to the formation of spinning fullerenes, whereas the formation of the initial corannulene, which with the addition of further pentagons and hexagons determines whether a spinning fullerene will be formed or may occur, a hemispherical cap may lift up and lead to the formation of an SWNT. This continuum elastic model for the nanoelasticity of carbon's shelled cages around the initial corannulenes has been developed to determine the criteria for the growth of these structures. Symmetry arguments have been used to explain the observed dominant growth of diameter 14Å, armchair SWNTs in sooting environments. Elaborate experiments will be needed to control the ingredients of the C vapour where fullerenes and nanotubes are formed to understand the full picture of the growth mechanisms. Simulations are the next best way to test such models and results of our initial investigations on the dynamics of closed cage formation with the addition of $C_1$, $C_2$ and $C_3$ indicate the requirement of the addition of $C_2$s for cage closure [30]. A final comment on future directions could be that since the present model uses continuum elasticity to probe the dynamical behaviour of molecular systems, a further project will be to look for a model that encompasses the quantum mechanical description of the nanoelastic behaviour of carbon's closed cages.